\documentclass[epj]{webofc}
\usepackage[varg]{txfonts}   
\usepackage{color,graphicx}
\usepackage{bm}
\usepackage{adjustbox}
\usepackage{adjustbox}
\usepackage{graphicx}
\usepackage{hyperref}
\hypersetup{colorlinks,%
            citecolor=black,%
            filecolor=black,%
            linkcolor=black,%
            urlcolor=black,%
            pdftex}
\usepackage{latexsym}

\newcommand{\ud}     {\mathrm{d}}

\newcommand{\Mev}    {\mathrm{MeV}}

\newcommand{\gevsq}  {\mathrm{GeV}^2}
\renewcommand{\Im}{\mathop{\mathrm{Im}}}
\renewcommand{\Re}{\mathop{\mathrm{Re}}}
\newcommand{\ceps}{\varepsilon}

\newcommand{\average}[1]{\langle{#1}\rangle}

\newcommand{\eq}[1]{Eq.(\ref{#1})}
\newcommand{\Eqs}[2]{Eqs.(\ref{#1}) and (\ref{#2})}
\newcommand{\eqs}[1]{Eqs.(\ref{#1})}

\begin{document}
\title{Nuclear Parton Distributions}
%
%

\author{%
S. A. Kulagin\inst{1}\fnsep\thanks{\email{kulagin@ms2.inr.ac.ru}}
}

\institute{Institute for Nuclear Research of the Russian Academy of Sciences, Moscow 117312, Russia
}

\abstract{%
We review a microscopic model of the nuclear parton distribution functions,
which accounts for a number of nuclear effects including
Fermi motion and nuclear binding, nuclear meson-exchange currents, off-shell corrections
to bound nucleon distributions and nuclear shadowing.
We also discuss applications of this model to a number of processes including
lepton-nucleus deep inelastic scattering, proton-nucleus Drell-Yan lepton pair production at Fermilab,
as well as $W^\pm$ and $Z^0$ boson production in proton-lead collisions at the LHC.
}
\maketitle

\section{Introduction}
\label{sec:intro}

The QCD factorization theorem \cite{Collins:1989gx} suggests that
the parton distribution functions (PDFs) are universal process-independent
characteristics of the target at high invariant momentum transfer $Q$.
PDFs determine the leading contributions
to the cross sections of various hard processes involving leptons and
hadrons. PDFs cannot be reliably calculated in modern QCD,
as they are driven by non-perturbative strong interactions,
and QCD-based phenomenology remains to be the primary source of information on PDFs.

The deep-inelastic scattering (DIS) experiments with nuclear targets
show significant nuclear effects
with a rate that is more than one order of magnitude larger than
the ratio of the nuclear
binding energy to the nucleon mass
(for a review see \cite{Arneodo:1992wf,Norton:2003cb}).
These observations rule out the naive picture of the nucleus as a system of quasi-free nucleons and
indicate that the nuclear environment plays an important role even
at energies and momenta much higher
than those involved in typical nuclear ground state processes.

The studies of the mechanisms responsible for observed effects
provide a link between particle and nuclear physics and help to
better understand both the intrinsic properties of the proton
as well as the properties of hadrons in nuclear environment.
A typical example in this context is the extraction of the $d$-quark PDF from the global fits
involving the proton and the deuteron data.
This procedure requires, in turn, a detailed
knowledge of nuclear effects in order to control
the corresponding systematic uncertainties.

A number of studies are available \cite{Eskola:2009uj,Hirai:2007sx,deFlorian:2011fp},
in which the nuclear parton distributions (NPDFs) are extracted
from global fits to nuclear high-energy data
using the empirical parameterizations of nuclear correction factors for each PDF.
Typically these studies involve a large number of free parameters
and can be biased by the chosen parameterizations of the Bjorken $x$
as well as the nuclear mass number $A$
dependencies of nuclear correction factors.
In this paper we follow a different approach and address the NPDFs
using the microscopic model developed in Ref.\cite{KP04}.
The model incorporates a number of nuclear corrections including the smearing with the
energy-momentum distribution of bound nucleons (Fermi motion and binding),
the off-shell correction to bound nucleon structure functions,
corrections from nuclear meson exchange currents (MEC)
and the propagation of
the hadronic component of the virtual intermediate boson in the nuclear environment.
In Ref.\cite{KP14} this approach was applied to compute individual NPDFs.
In Sec.\ref{sec:npdf} we briefly review the approach in context of nuclear DIS.
The applications to the nuclear Drell-Yan (DY)
reaction in the context of the measurement of nuclear antiquark distributions
in Fermilab E772 experiment is discussed in Sec.\ref{sec:dy}.
In Sec.\ref{sec:wz} we review application of this approach to study
the $W$ and $Z$ boson production in p+Pb collisions at LHC.

\section{Nuclear DIS and PDFs}
\label{sec:npdf}

In a reference frame in which the hadron has a high (infinite) momentum,
a PDF is the momentum distribution of the corresponding parton in a hadron in the units of hadron's
momentum. However, the interpretation of PDFs is somewhat more complicated in the nucleus rest frame.
We recall that in the target rest frame the characteristic DIS time (or longitudinal distance)
can be estimated as  $L\sim (Mx)^{-1}$, where $M$ is the nucleon mass and
$x$ the Bjorken scaling variable \cite{Ioffe:1985ep}.
Using $L$ we identify two different kinematical regions for nuclear effects.
In the region $L<d$, where $d$ is the average distance between bound nucleons in a nucleus,
i.e. for large values of $x>0.1$,
the nuclear DIS cross sections can be approximated by incoherent scattering
from bound protons and neutrons.
The dominant nuclear corrections in this region are due to energy-momentum
distribution of bound nucleons (nuclear Fermi motion and binding
\cite{West74,Bodek:1980ar,Akulinichev:1986gt,Akulinichev:1985ij})
and also off-shell correction to bound nucleon structure functions \cite{Kulagin:1994fz}.
In the region of small $x<0.1$ one has to account for
a scattering corrections from nuclear meson fields
\cite{LlewellynSmith:1983vzz,Ericson:1983um}.
Also in the region $L\gg d$ (or $x\ll0.1$)  the propagation effects of
virtual quark-gluon states in the nuclear environment results in essential corrections
to the impulse approximation.
For the leading contributions to the DIS structure functions
the interference of multiple scattering terms results in a negative
correction (nuclear shadowing, for a review see, e.g., \cite{Piller:1999wx}).

We use the notation $q_{a/h}(x,Q^2)$ for the distribution of (anti)quarks of the type
$a=u,\bar u, d,\bar d,\ldots$ in a hadron $h$.
Following Ref.\cite{KP04,KP14} we write the nuclear PDF $q_{a/A}$
as (for brevity, we suppress explicit dependencies on $x$ and $Q^2$)
\begin{equation}
\label{npdf}
q_{a/A} = q_{a/A}^\mathrm{IA} + q_{a/A}^\mathrm{coh} + q_{a/A}^\mathrm{MEC},
\end{equation}
where the first term on the right side is the contribution from
bound protons and neutrons in the impulse approximation, and the other
terms are the corrections to the impulse approximation due to
coherent nuclear interactions of the hadronic component of the virtual photon
and to nuclear meson exchange currents, respectively.
These contributions are reviewed below.

\subsection{Impulse approximation}
\label{sec:ia}

In the impulse approximation the nuclear PDFs can be written as
a convolution of the proton (neutron) distribution of a nucleus with the
corresponding parton distribution of a bound proton (neutron) \cite{Kulagin:1994fz,KP04}:
\begin{align}
\label{eq:IA}
q_{a/A}^\mathrm{IA} &= \sum_{\tau=p,n} f_{\tau/A} \otimes q_{a/\tau}
= \sum_{\tau=p,n} \int_{x<y}\frac{\ud y\ud p^2}{y} f_{\tau/A}(y,p^2) q_{a/\tau}(x/y,Q^2,p^2).
\end{align}
The nuclear convolution is an integration over both the nucleon
light-cone momentum $y$ and the nucleon invariant mass (virtuality) $p^2$,
since PDFs of an off-shell nucleon generally depend on its virtuality~\cite{Kulagin:1994fz}.
The proton (neutron) distribution function in a nucleus is given in terms of
the nuclear spectral function $\mathcal{P}$
(for brevity, we drop subscripts identifying the proton and the neutron distributions)
\begin{align}
\label{eq:f}
f(y,p^2)= \int [\ud k] \left(1+\frac{k_z}{M}\right)\mathcal{P}(\bm{k},\ceps)
\delta\left(y - \frac{k_0+k_z}{M}\right) \delta\left(p^2 - k^2\right),
\end{align}
where the integration runs over the
nucleon four-momentum $k=(M+\ceps,\bm k)$  and $[\ud k]=\ud k_0 \ud\bm k/(2\pi)^4$ and
$k^2=k_0^2-{\bm k}^2$ is the invariant mass of the off-shell nucleon.
The coordinate system is such that the momentum transfer is antiparallel to the $z$ axis.
The Bjorken variable of the nucleus is $x=Q^2/2Mq_0$
with $q_0$ the energy transfer in the target rest frame.

We refer \eq{eq:IA} as a generalized impulse approximation because it accounts
for the dependence of the nucleon PDF on the nucleon invariant mass $p^2$,
which is usually not addressed in various versions of impulse approximation.
A similar nuclear convolution equation but without off-shell effect
was earlier discussed in \cite{Kulagin:1989mu,Akulinichev:1986gt,Akulinichev:1985ij}.
Note that \Eqs{eq:IA}{eq:f} were obtained starting from a Lorentz-covariant approach and
using a systematic expansion of matrix elements in series of the small parameters
$\bm p/M$ and $\ceps/M$, keeping terms of the order $\bm p^2/M^2$ and $\ceps/M$
\cite{Kulagin:1989mu,Kulagin:1994fz,KP04}.
To this order a generalized nucleon distribution function (\ref{eq:f})
is given in terms of nonrelativistic spectral function
\begin{equation}\label{eq:P}
\mathcal P(\bm k,\ceps) = \int\ud t\, e^{i\ceps t} \average{\psi^\dagger(\bm k,t)\psi(\bm k,0)},
\end{equation}
where $\psi(\bm k,t)$ is the nonrelativistic nucleon operator in the momentum-time representation
(for more details see Ref.\cite{KP04}). The spectral function describes the energy-momentum
distribution of bound nucleons.
Note that $\ceps$ in \eq{eq:P} includes the recoil kinetic energy of the residual system of $A-1$ nucleons,
as it can be seen after inserting a complete set of states and integrating over the time.
\footnote{Note that for the deuteron the spectral function is given in terms of the deuteron wave
function as
$\mathcal P_d=2\pi\left|\Psi_d(\bm k)\right|^2\delta(\ceps-\ceps_d -\bm k^2/2M)$ with
$\ceps_d=-2.22\,\Mev$ the deuteron binding energy \cite{KP04}.
}

The proton (neutron) spectral function is normalized to the proton $Z$ (neutron $N$) number, and
using \eq{eq:f} we explicitly verify that the proton (neutron) distribution function is normalized
as
\begin{equation}\label{norm}
\int\ud y \ud p^2 f_{p(n)/A}(y,p^2) = Z (N),
\end{equation}
where the integral is taken over all possible light-cone momenta $y$ and the nucleon virtuality $p^2$.
Note that the term $k_z/M$ in \eq{eq:f} gives vanishing contribution to the normalization (\ref{norm})
due to symmetry reason.
The distribution function (\ref{eq:f}) is independent of $Q^2$ in the Bjorken limit
and the $Q^2$ evolution of the NPDFs in the impulse approximation is governed by the evolution
of the PDFs of the corresponding nuclear constituents.
For the discussion of power  corrections to the nuclear convolution (\ref{eq:IA})
we refer to Ref.\cite{KP04,Kulagin:1997vv,Kulagin:2000yw}
(see also \cite{Kulagin:2008fm} for spin-dependent DIS).

The first moment of the nucleon distribution in $y$ gives the fraction of the nuclear
light-cone momentum carried by bound nucleons.
The first moment in $p^2$ gives average nucleon virtuality.
It will be convenient to discuss dimensionless virtuality $v=(p^2-M^2)/M^2$. We have
\begin{align}
\label{eq:avy}
\average{y}_N &=\frac 1A \int\ud y\ud p^2 y f(y,p^2)=1+\frac{\average{\ceps}+\tfrac23\average{T}}{M},
\\
\label{eq:avv}
\average{v}_N &=\frac 1A \int\ud y\ud p^2 v f(y,p^2)=2\frac{\average{\ceps}-\average{T}}{M},
\end{align}
where we sum over protons and neutrons and
$\average{\ceps}$ and $\average{T}=\average{\bm k^2}/2M$
are the nucleon energy and kinetic energy integrated with the nuclear spectral function (\ref{eq:P})
per one nucleon.
Note that $\average{\ceps}-\average{T}=\average{V}$ is the average nuclear potential energy
per one nucleon.
For the deuteron we can evaluate the moments in \Eqs{eq:avy}{eq:avv}
using the deuteron wave function. For the Paris wave function we have
$\average{y}_N=0.994$ and $\average{v}_N=-0.044$.
For $^{208}$Pb, using the nuclear spectral function from \cite{KP04},
we have  $\average{y}_N=0.963$ and $\average{v}_N=-0.197$.
Note that $\average{y}_N<1$ and $\average{v}_N<0$ because of nuclear binding.
The inequality $\average{y}_N<1$ indicates that the bound nucleons carry
only a part of the full nuclear light-cone momentum. In this approach
the missing nuclear light-cone momentum is carried by the nuclear meson
degrees of freedom which are responsible for nuclear binding,
as discussed in Sec.\ref{sec:mec}.

In terms of the variable $y$,
the distribution function $f(y,p^2)$ is peaked about $y=\average{y}_N$ with
a width $\sim p_F/M$ with $p_F$ the nuclear Fermi momentum.
In terms of dimensionless virtuality $v$,
the distribution $f(y,p^2)$ is located in a narrow region
about $v=\average{v}$.

Note that \eq{eq:IA} predicts a finite nuclear PDF in the region
$x>1$, because of high momenta $p_z>p_F$.
The rate of NPDF in this region is driven by the high-momentum part of the nuclear momentum distribution.%
\footnote{Allowed region of nuclear Bjorken variable is
$0<x<M_A/M$, where $M_A$ is the mass of the nucleus.
}
In the applications discussed below
we use a model spectral function, which includes both a mean field contribution
dominant at low energy and momentum,
and a high-momentum and high-energy component related to
short range nucleon-nucleon correlations (SRC) \cite{KP04}.
In order to address the effect of nonisoscalarity (nuclear neutron excess)
we use the isoscalar $\mathcal P_0=\mathcal P_{p}+\mathcal P_{n}$
and the isovector $\mathcal P_1=\mathcal P_{p}-\mathcal P_{n}$
combinations of the nuclear spectral function.
We assume the SRC contribution to be similar for the proton and the neutron
nuclear distribution such that the SRC term only contributes to
the isoscalar spectral function
$\mathcal P_0$
and that it cancels out in the isovector combination
$\mathcal P_1$.
This behavior is supported by the observation of the dominance of $pn$ SRC pairs
in nucleon knock-out experiments \cite{Subedi:2008zz}.
The isovector spectral function $\mathcal P_1$ is therefore
calculated as the difference between the mean-field contributions
to the proton and the neutron spectral functions and is proportional
to the proton-neutron asymmetry $\beta=(Z-N)/A$ \cite{KP04}.

In off-mass-shell region the PDFs explicitly depend on the nucleon
virtuality as indicated in \eq{eq:IA}.
In order to address this effect we note that
on average the nucleon virtuality $v$ is small. We then expand
the off-shell PDF $q(x,Q^2,p^2)$ in \eq{eq:IA} about the on-shell
value $p^2=M^2$ in series in $v$ \cite{Kulagin:1994fz}. Keeping the leading terms we
have \cite{KP04,KP14}:
%
\label{OS:def}
\begin{align}
\label{pdf:os}
q(x,Q^2,p^2) &\approx q(x,Q^2)(1+\delta f(x,Q^2) v),
\\
\delta f(x,Q^2) &= \partial \ln q(x,Q^2,p^2)/\partial \ln p^2 ,
\label{deltaf:def}
\end{align}
where the derivative is evaluated at $p^2=M^2$ and
$q(x,Q^2)$ is the PDF of the on-shell nucleon.
The function $\delta f$ describes the
relative off-shell modification of the nucleon PDF.
This special nucleon structure function does not contribute to the cross section of the
physical nucleon, but it is relevant only for the bound nucleon and describes its
response to the interaction in a nucleus.
The function $\delta f$ was studied phenomenologically%
\footnote{%
Nuclear DIS with off-shell effect was discussed in a number
of different approaches \cite{Gross:1991pi,Kulagin:1994fz,Bondarenko:2002zz,Alekhin:2003qq}.
}
in an analysis
of data on the nuclear DIS \cite{KP04} (see Sec.\ref{sec:dat}).
This analysis suggests a common off-shell function for the quark and antiquark
distributions, independent of $Q^2$ and  the parton type.
This observation is supported by a recent QCD global analysis of proton and deuteron data \cite{AKP16},
as well as by the studies of the nuclear DY production
of muon pair in Sec.\ref{sec:dy} and the $W/Z$ boson production in p+Pb
collisions in Sec.\ref{sec:wz}.

\subsection{Nuclear MEC correction}
\label{sec:mec}

As it was noted in Sec.\ref{sec:ia} the nuclear binding effect causes
$\average{y}_N<1$. Apparently the missing nuclear
light-cone momentum should be carried by the fields responsible for nuclear
binding \cite{LlewellynSmith:1983vzz}.
The correction from the nuclear virtual meson cloud
was discussed in the context of the nuclear EMC effect in a number of papers
\cite{LlewellynSmith:1983vzz,Ericson:1983um,Friman:1983rt,Sapershtein:1985pa,Jung:1990pu,Kaptari:1989un,Marco:1995vb}.
These studies indicate an enhancement of the nuclear structure functions
in the region $0.05<x<0.15$ due to nuclear meson cloud.
However,  the results of specific calculation are somewhat uncertain due to
sensitivity to the details of the pion-nucleon form-factor,
the treatment of the particle-hole nuclear excitations
in intermediate state as well as the $\Delta$-resonance degrees of freedom.

Here we briefly review the approach followed in Ref.\cite{Kulagin:1989mu,KP04,KP14}
and consider constraints on the common nuclear meson cloud
coming from the nuclear light-cone momentum sum rule
together with the meson contribution to the nuclear potential energy.
We write the MEC correction in a convolution form similar to \eq{eq:IA}:
\begin{equation}\label{eq:mec}
q_{a/A}^\mathrm{MEC}=\!\sum_{m=\pi,\rho,\ldots} f_{m/A} \otimes q_{a/m}
\end{equation}
where $f_{m/A}(y,k^2)$ is the nuclear distribution function of a meson type $m$
over light-cone momentum $y$ and invariant mass  $k^2$ and
$q_{a/m}$ is corresponding PDF in a virtual meson $m$ with invariant mass $k^2$.
The meson distribution function $f_m$ is driven by the imaginary part
of the full meson propagator in a nucleus.


For simplicity we will  discuss the isoscalar part of $q_{a/A}^\mathrm{MEC}$
assuming the meson PDFs $q_{a/m}$ to be independent from the meson type and equal to the
corresponding pion PDF.
We also neglect possible off-shell dependence of $q_{a/m}$.
This will allow us to discuss the
light-cone distribution function corresponding to the sum over all mesons
$f_M(y)=\sum_m\int\ud k^2 f_m(y,k^2)$.

The nuclear light-cone momentum is shared between the nucleons and mesons
and the corresponding balance equation can be written as \cite{Kulagin:1989mu}
\begin{equation}\label{balance:eq}
\average{y}_M + \average{y}_N = \frac{M_A}{AM},
\end{equation}
Note the ratio of the nucleus mass $M_A=A(M+\ceps_B)$,
where $\ceps_B$ is the nuclear binding energy per nucleon, and the nucleon
mass $M$ in \eq{balance:eq}. This ratio appears because of the definition of
the light-cone momentum $y$ and the nuclear Bjorken variable $x$ involves
the nucleon mass $M$.

To further constrain the meson distribution,
we consider a model nuclear Hamiltonian in effective meson-nucleon theory.
Using equations of motion for interacting mesons and nucleons
it was possible to obtain a relation \cite{Kulagin:1989mu,KP04,KP14}
\begin{equation}\label{eq:m2}
3\average{y}_M + \frac{m^2_M}{M^2}\average{y^{-1}}_M = -\frac{\average{V}}{M},
\end{equation}
where $\average{y}_M$ and $\average{y^{-1}}_M$ are the moments
of the nuclear meson distribution function, $m_M^2$ is an average meson
mass squared, and $\average{V}$ is the average nuclear potential energy
(see \eq{eq:avv} and the discussion thereafter).
Note that \eq{eq:m2} was obtained in static approximation neglecting
the contribution from the terms with time derivatives of meson fields
(this is justified by the fact that a typical energy transfer in the
 nucleon--nucleon interaction in the nuclear ground state is much smaller
 than corresponding momentum transfer).
If only the pion contribution is considered, then $m_M^2=m_\pi^2$ and
$\average{V}$ corresponds to the pion contribution to the nuclear potential energy.

It is important to note that \Eqs{balance:eq}{eq:m2} allow us to overall constrain
the nuclear meson distribution function in terms of average nucleon separation and
kinetic energy $\average{\ceps}$ and $\average{T}$.
In particular, for the deuteron with the Paris wave function
we have $\average{y}_M=0.0045$ and $\average{y^{-1}}_M=0.390$,
while for the $^{208}$Pb nucleus
we have 0.029 and 0.543, respectively.

In applications discussed below we consider a model meson distribution function
which obeys the constraints discussed above.
Following Ref.\cite{Kulagin:1989mu,KP04} we assume $f_{M/A}(y)$ to scale as $y$ as $y\to 0$.
We also recall that large values of $y$ are driven by
configurations with $k_0$ and $k_z$ comparable to or exceeding the nucleon mass $M$.
We assume that such configurations are suppressed in nonrelativistic nuclei and
consider the meson
distribution in the region $0<y<1$ and use the following model
\begin{equation}\label{fpi:model}
f_{M/A}(y) =c\,y(1-y)^n.
\end{equation}
The parameters $c$ and $n$ depend on the specific nucleus and
fixed from $\average{y}_M$ and $\average{y^{-1}}_M$ calculated by \Eqs{balance:eq}{eq:m2}.

\subsection{Correction from nuclear coherent processes}
\label{sec:coh}

In the region of small $x$ the DIS correlation length $L$ exceeds
typical distances between bound nucleons and intermediate quark-gluon states
may develop multiple scattering interactions while propagating in nuclear environment.
To the leading order this effect results in
a negative correction known as the nuclear shadowing effect
(for a review, see Ref.\cite{Piller:1999wx}).

In order to address this effect, in Ref.\cite{KP04} we assume that
the set of intermediate hadronic states can be approximated by a single
effective state and describe its interaction with the nucleon
by an effective scattering amplitude $a$.
The rate of the contribution to DIS cross section is driven by $\Im a$.
This amplitude depends on the type of the PDF and differ for the proton and the neutron.

It is convenient to discuss nuclear effects in terms of the ratio
$R_{i/A}= q_{i/A}/(Zq_{i/p}+Nq_{i/n})$
for the given PDF type $i$.
Using the optical theorem this ratio can be written in terms of effective cross sections,
or the imaginary part of the effective amplitudes in the forward direction
\begin{equation}
\delta_\mathrm{coh} R_{i/A} = \Im \mathcal A_i/(Z\Im a_{ip}+N\Im a_{in}),
\end{equation}
where
$\mathcal A_i$ is the nuclear amplitude which describes
propagation of intermediate states corresponding to PDF of type $i$ in a nucleus,
and $a_{ip}$ and $a_{in}$ are corresponding proton and neutron amplitudes, respectively.
We compute the amplitude $\mathcal A$ in terms of the Glauber-Gribov multiple-scattering series
\cite{Glauber:1970jm,Gribov:1968gs}.
Note that the multiple-scattering series start from the double-scattering term,
as the single-scattering term is already accounted in the impulse approximation of \eq{eq:IA}.
For more detail discussion of $\mathcal A$ see Ref.\cite{KP04,KP14}.

To have a closer link to the DIS structure functions,
it is convenient to discuss
the combinations of PDFs with definite $C$-parity, $q_i^\pm = q_i \pm \bar q_i$.
Let $a_{ip}^\pm$ and $a_{in}^\pm$ be the corresponding proton and neutron effective
amplitudes.
For the $u$ and $d$ (anti)quark distributions we also consider
the combinations with definite isospin $I=0,1$ and $C$ parity $q_I^C$, i.e.
$q_0^\pm=u^\pm+d^\pm$ and $q_1^\pm=u^\pm-d^\pm$.

We separate the isoscalar and the isovector contributions in $\mathcal A$
assuming the isospin symmetry for the scattering off protons and neutrons, i.e.
$a_{up}=a_{dn}$ and $a_{dp}=a_{un}$, and write the amplitudes as
$a_{up}=a_0+\tfrac12 a_1$ and $a_{dp}=a_0-\tfrac12 a_1$,
where $a_0$ and $a_1$ are the isoscalar and isovector amplitudes, respectively.
To the first order in $\beta=(Z-N)/A$ we have \cite{KP07}
\begin{equation}\label{A:ud}
\mathcal{A}_{u,d} = \mathcal{A}(a_0) \pm
	\frac{\beta}{2} a_1 \mathcal{A}'(a_0) ,
\end{equation}
where the sign $+-$ should be taken for the $u$ and $d$ quark, respectively,
and $\mathcal A'=\partial \mathcal{A}/\partial a$.
The first and the second terms in  \eq{A:ud} drive the corrections to the
isoscalar $q_0$ and the isovector $q_1$ quark distribution, respectively.

Let us consider first the isoscalar $I=0$ case.
For the nuclear corrections for the $C$-even
and $C$-odd quark distributions we have
\begin{subequations}\label{npdf:coh:0}
\begin{align}
\label{coh:0pl}
\delta {R}^{+}_0 &= \Im \mathcal A(a_0^+)/(A \Im a_0^+) ,
\\
\label{coh:0mn}
\delta {R}^{-}_0 &= \Im [a_0^{-} \mathcal A'(a_0^+)]/(A \Im a_0^{-}) ,
\end{align}
\end{subequations}
where $a_0^\pm$ are the $I^C=0^\pm$ amplitudes.
We note that \eqs{npdf:coh:0} are obtained by treating the $C$-odd amplitude as a small parameter and
expanding the difference between the quark and antiquark nuclear
amplitudes in series of $a_0^-$ to the order $(a_0^-)^2$ \cite{KP07}.
The effective expansion parameter in \eqs{npdf:coh:0} is the ratio of the amplitudes
$a_0^-/a_0^+$. The smallness of this parameter can be justified
within the Regge pole model of high-energy scattering amplitudes.

The nuclear corrections in the isovector term $I=1$ can be calculated
similarly to the isoscalar case discussed above.
We consider $1^\pm$ channels and expand the corresponding nuclear amplitude
in series of $a_0^-$. To the leading order we have
\begin{align}\label{npdf:coh:1}
\delta {R}^{\pm}_1 &= \beta
	\Im \left[	a_1^{\pm} {\mathcal A}'(a_0^+) \right]/(A \Im a_1^\pm) ,
\end{align}
where the superscript $+$ and $-$ corresponds to the channel $1^+$ and $1^-$, respectively.
In the derivation of \eq{npdf:coh:1} we drop the terms of order $a_1^\pm a_0^-$.
Note in this context that the effective amplitudes  $a_1^\pm$ and $a_0^-$ are
generally significantly smaller than the leading amplitude $a_0^+$,
which  drives multiple scattering corrections
for all PDFs, as it can be seen from \Eqs{npdf:coh:0}{npdf:coh:1}.
Furthermore, in the considered approximation
the ratio $\delta R_0^-$ as well as the ratios $\delta R_1^\pm$
are independent of the effective cross sections in the corresponding channels and depend
only on  $\alpha=\Re a/\Im a$ of corresponding amplitude.

The individual corrections for $u$ and $d$ quarks and antiquarks can be derived
from $\delta R_{0,1}^\pm$ ratios. For more detail see Ref.\cite{KP14}.

\subsection{Discussion and comparison with data}
\label{sec:dat}

A detailed analysis of data on the ratios of DIS structure
functions $ R(A/A')=F_2^A/F_2^{A'}$ for different nuclei
was carried out in Ref.\cite{KP04} in the context of the described model.
The analysis included data with $Q^2\ge1\,\gevsq$ for the full region of Bjorken $x$
from CERN, FNAL and SLAC available before 1997 (see Table~1 in Ref.\cite{KP04}).
The ratio $R(A/A')$ was computed including nuclear corrections discussed above.
Also the target mass \cite{TMC} and the higher-twist corrections \cite{Alekhin:2007fh}
were applied to the structure functions.
The predictions were then compared with data by evaluating $\chi^2$
as discussed in Ref.\cite{KP04}.

\begin{figure*}[p]
\begin{center}
\includegraphics[width=\textwidth,height=0.91\textheight]{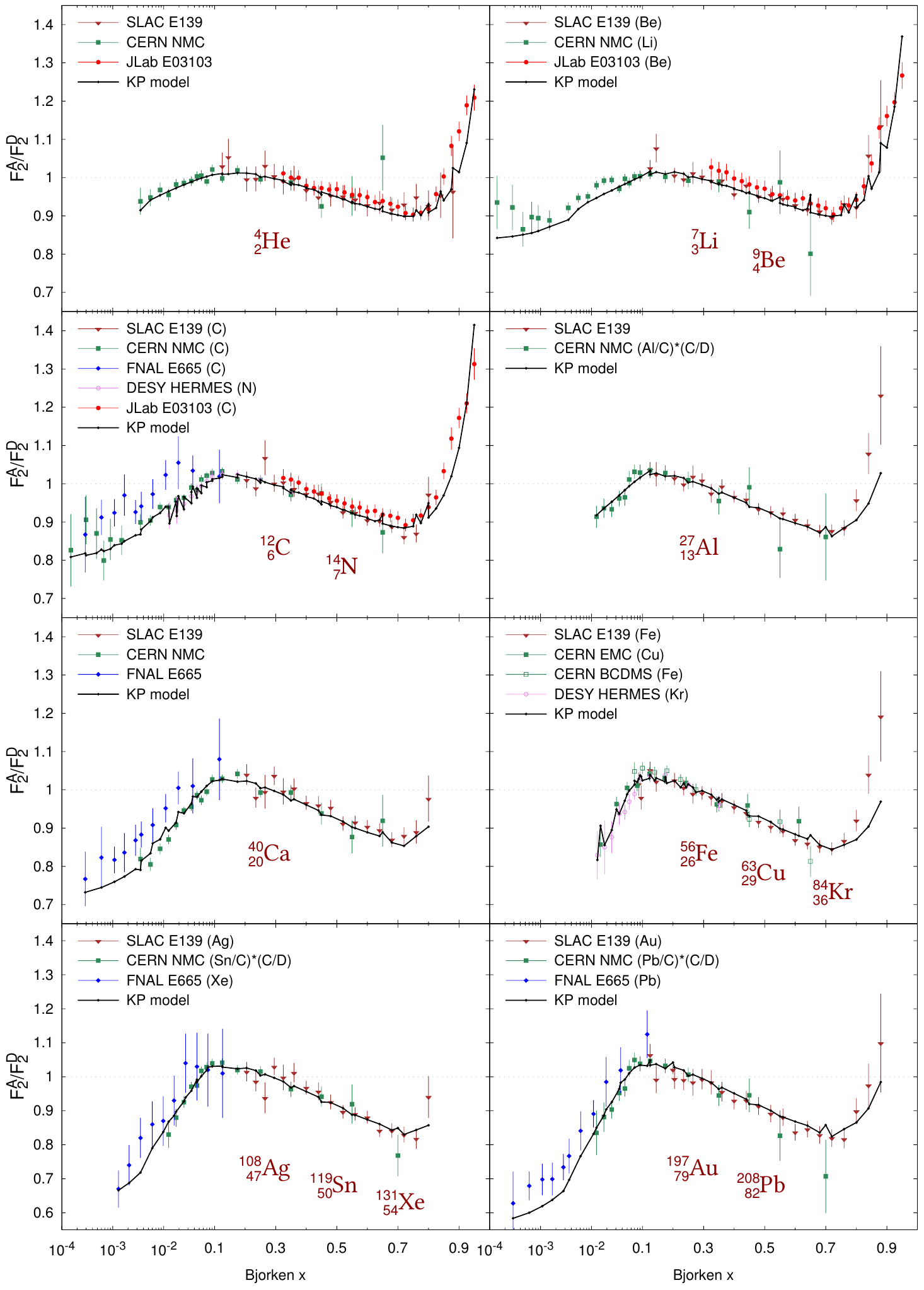}
\end{center}
\caption{\label{fig:emc-effect}
Summary of data on the ratio $F_2^A/F_2^D$ for various nuclei
from the measurements at CERN \cite{EMC,NMClic,NMChe,NMCpb,BCDMS},
SLAC \cite{E139},
FNAL \cite{E665},
JLab \cite{JLAB}
and DESY \cite{HERMES}. The error bars represent the statistical and systematic uncertainties
added in quadrature.
The dots connected by the solid line are the predictions of Ref.\cite{KP04}
computed for the published values of $(x,Q^2)$ of each data point
(the wiggles are caused by $Q^2$ dependence and different values of $Q^2$ from different experiments).
We use a logarithmic scale for $x<0.1$ and a linear scale for $x>0.1$ for a
better display of both the small $x$ and the large $x$ regions.}
\end{figure*}

The function $\delta f$, which determines off-shell correction (see \eq{deltaf:def}),
and also the effective amplitude $a_0^+$, which drives coherent nuclear corrections
for the structure function $F_2$ at small $x$,
were determined phenomenologically from this analysis.
The function $\delta f$ was assumed universal,
i.e. same for quark and antiquark PDFs independent of flavor, and
also independent of scale $Q^2$, $\delta f=\delta f(x)$. We model $\delta f(x)$
as a third-order polynomial and extract parameters from analysis of $ R(A/A')$.
The effective amplitude $a_0^+$ is the subject of several constraints.
In the region of low $Q^2<1\,\gevsq$ the amplitude $a_0^+$ is constrained
by vector meson dominance model \cite{Bauer:1977iq}.%
\footnote{%
Although this region was excluded from our fit,
it was used to verify the predictions of Ref.\cite{KP04}.
}
Also in the region  $Q^2> 10\,\gevsq$ the amplitude $a_0^+$ and
the function $\delta f$ are linked by the normalization condition on the
nuclear valence quark distribution. The latter is used to determine the
leading twist part of the amplitude $a_0^+$ \cite{KP04,KP14}.
The available nuclear DIS data at small $x$ constrains
the effective amplitude $a_0^+$ in a region which is transitional
between fully nonperturbative region of $Q^2<1\,\gevsq$ and
the region $Q^2>10\,\gevsq$, which is driven by the leading twist contribution \cite{KP04}.

The results reported in Ref.\cite{KP04} show an
accurate description of the measured dependencies on
$x$, $Q^2$ and the nuclear mass number $A$ in the full kinematical region of data.
The predictions of Ref.\cite{KP04} were further verified~\cite{KP10}
with the recent nuclear DIS data from HERA \cite{HERMES} and  JLab \cite{JLAB}.
Figure~\ref{fig:emc-effect} summarizes the data on $R(A/D)$
from $^4$He to ${}^{208}$Pb together with the model calculations.


We would like to remark that in some cases the
data points from different experiments are not fully consistent. In
particular, the central points of ${}^{12}$C/D and ${}^{40}$Ca/D ratios from
E665 experiment \cite{E665} at low $x$ are systematically above the
corresponding NMC measurements, which have smaller uncertainties.
Similarly, a normalization problem could be present for the E665 ${}^{208}$Pb/D
data.
However, the double ratios (${}^{40}$Ca/D)/(${}^{12}$C/D) and
(${}^{208}$Pb/D)/(${}^{12}$C/D) of the E665 measurement
are in good agreement with the NMC data \cite{NMCpb}
as well as with our predictions.
Also the central points of $^4$He/D, $^9$Be/D and $^{12}$C/D ratios reported in
Ref.\cite{JLAB} are systematically above the corresponding E139 \cite{E139}
and NMC \cite{NMClic,NMChe} results for $x>0.3$.%
\footnote{Similarly, about 3\% mismatch in the normalization of data points
from HERMES and JLab measurements is present for $^3$He/D ratio.}
However, the slopes of the measured ratios for $x>0.2$
seems to be in a good agreement for all experiments.
We also comment that the analysis of Ref.\cite{KP10} indicates
that a common renormalization factor of 0.98 applied to the data points of Ref.\cite{JLAB}
leads to a perfect statistical agreement of the discussed data sets.

\begin{figure*}[htb]
\begin{center}
\includegraphics[width=\textwidth]{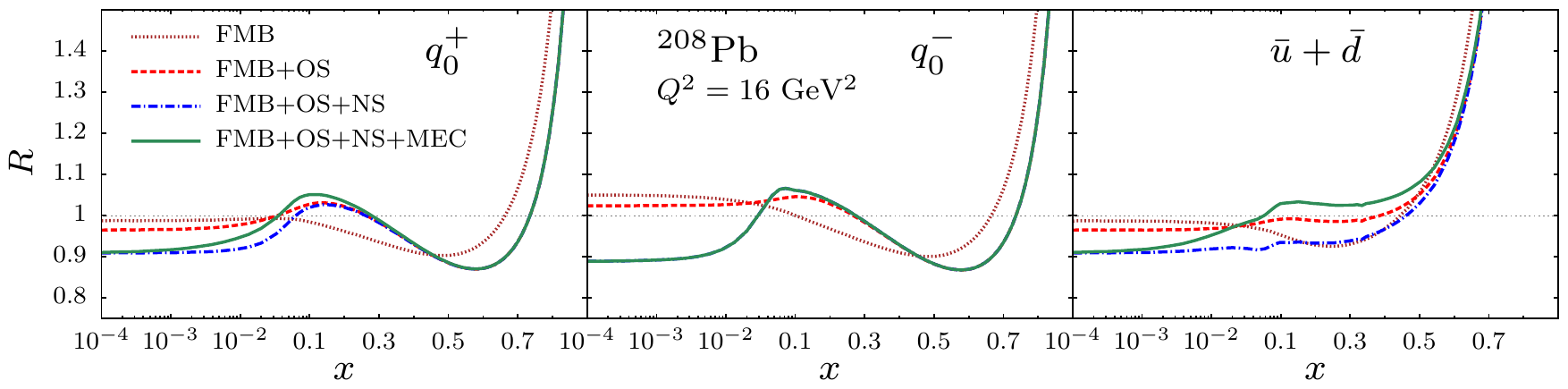}
\end{center}
\caption{\label{fig:npdf}
The ratio $R_i=q_{i/A}/(Zq_{i/p}+Nq_{i/n})$ computed for various combinations
of parton distributions for the lead nucleus at fixed $Q^2=16\,\gevsq$.
The left panel shows different nuclear corrections for $q_0^+=u+d+\bar u+\bar d$,
the middle panel is for $q_0^-=u+d-\bar u-\bar d$,
while the right panel is for antiquark distribution $\bar u+\bar d$.
}
\end{figure*}

In Fig.\ref{fig:npdf} we show nuclear effects on different PDFs computed
for $^{208}$Pb at $Q^2=16\,\gevsq$ following Ref.\cite{KP14}.
The labels on the curves show different nuclear corrections
included in turn:
smearing with nuclear spectral function (Fermi motion and binding, or FMB),
off-shell correction (OS),
nuclear shadowing (NS),
meson exchange currents (MEC).

Note that different nuclear effects outlined above are relevant in
different regions of Bjorken $x$, as it can be seen from Fig.\ref{fig:npdf}.
In the region of large $x>0.2$ the leading
correction is due to smearing with the nuclear spectral function.
The off-shell correction is also relevant in this region. The nuclear MEC correction is
relevant for $x<0.2$ while at small $x\ll 0.1$ the nuclear effects are
dominated by multiple scattering of virtually produced quark-gluon
intermediate states.
Note also an interplay between different nuclear corrections
on the antiquark distribution leading to a strong cancelation of
nuclear effects in an intermediate region of $x$.
The implication of this observation will be discussed in more detail
in Sec.\ref{sec:dy} in the context of the measurement of nuclear sea-quark
distribution in the Drell-Yan experiments.

We also remark that different nuclear effects are related through
a number of sum rules which are driven by global symmetries.
In particular, the conservation of the valence quark number
links together the off-shell and the shadowing corrections
to the valence quark distributions. The normalization
conditions for the nuclear isoscalar $q_0^-$ and the isovector $q_1^-$
valence quark distributions were used in this analysis to constrain
the unknown amplitudes $a_0^-$ and $a_1^-$
controlling the nuclear correction in the small $x$ region~\cite{KP14}.
The conservation of the nuclear light-cone momentum causes the corresponding sum rule
at two different levels.
At the hadronic level, the nuclear light-cone momentum is shared
between nucleons and mesons (see \eq{balance:eq}), which allows us to
constrain the MEC correction to NPDFs.
At the partonic level, the light-cone momentum is balanced between quarks,
antiquarks and gluons.
The study of different contributions to the light-cone momentum sum rule
can provide insights on modification of gluon distribution in nuclei.

\section{Nuclear Drell-Yan process}
\label{sec:dy}

The reaction of muon pair production in hadron-hadron collisions (Drell-Yan process)
is an important source of information on the proton, pion and nuclear PDFs \cite{Peng:2014hta}.
In the context of NPDFs, the use of DY data in combination with DIS data allows a separation
of the nuclear valence and sea quark distributions.
In the DY reaction with the proton beam by tuning the kinematics of the muon pair
one can select a region in which the DY cross sections are driven by annihilation of valence
quarks in the beam and antiquarks in the target. Then the ratio of the p+A DY cross sections
off different nuclear targets provide
a tool to measure the nuclear dependence of antiquark PDFs
\begin{equation}\label{eq:dy_r}
\frac{\sigma^{DY}_A}{\sigma^{DY}_B}
	\approx
	\frac{\bar u_A(x_T,Q^2) + \bar d_A(x_T,Q^2)}
	{\bar u_B(x_T,Q^2) + \bar d_B(x_T,Q^2)} ,
\end{equation}
where $x_T$ is the Bjorken variable of a nuclear target
and $Q$ is the mass of the lepton pair.

\begin{figure*}[hbt]
\begin{center}
\includegraphics[width=0.9\textwidth]{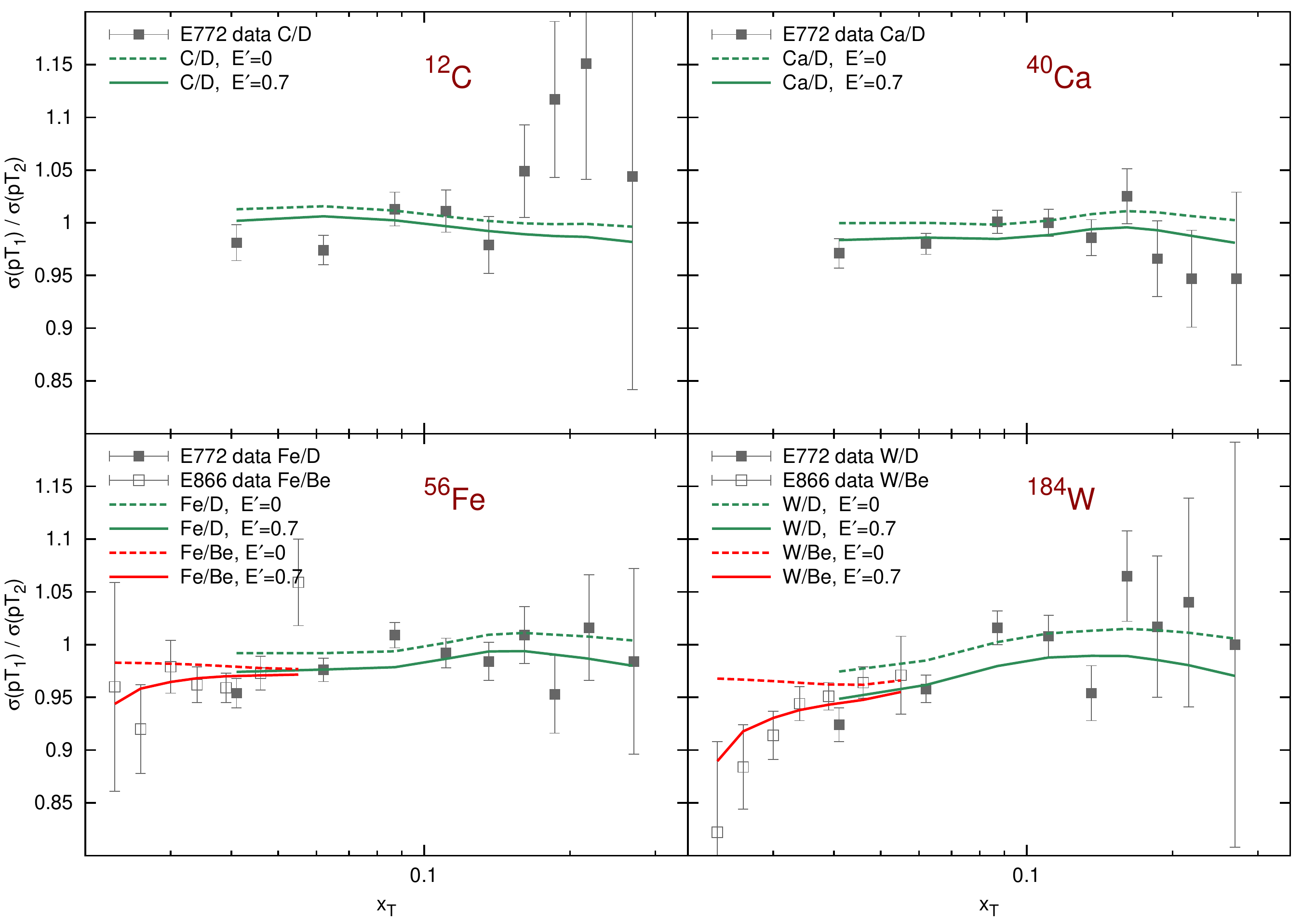}
\end{center}
\caption{%
Ratio of the DY reaction cross sections for different nuclei vs. $x_T$.
Data points are from the E772 experiment~\cite{E772} (full squares)
and E866 experiment~\cite{E866} (open squares).
Note that the ratio is normalized to one bound
nucleon and taken relative to the deuteron for E772
and $^9$Be for E866.
The curves are the predictions of Ref.\cite{KP14} for the deuteron (green) and beryllium (red) ratios
with (full line) and without (dashed line) the projectile energy loss effect
(see text and legend).
\label{fig:dy}}
\end{figure*}

This ratio  was measured in the experiments E772~\cite{E772} and E866~\cite{E866}
with the proton beam momentum 800 GeV/c at Fermilab in the region
$x_T < 0.15$ for a number of nuclear targets.
Note that the scale $Q^2$ is determined by the dimuon mass and $Q^2>16\,\gevsq$ in \cite{E772},
which is significantly higher than the corresponding scale from the fixed-target nuclear DIS
in the sea region.
In contrast to DIS, the data on DY nuclear ratios show no antishadowing
(i.e. enhancement of nuclear antiquark distributions)
at $x_T\sim 0.1$ that was a long standing puzzle since the nuclear
binding should result in an
excess of nuclear mesons, which is expected to produce a marked enhancement in the nuclear
anti-quark distributions \cite{Bickerstaff:1985ax}.

In Fig.\ref{fig:dy} we show the data along with our predictions on the ratios
of the DY cross sections \cite{KP14}.
Note that, as it can be seen from Fig.\ref{fig:npdf}, various nuclear corrections
to the antiquark distribution tend to cancel each other in the region
$0.1<x_T<0.4$, in agreement with data on the nuclear DY process.
We also remark that nuclear dependence of the DY process comes from two different sources:
(i) the nuclear effects on the bound nucleon PDFs,
and
(ii) the initial state interaction of the projectile parton in nuclear environment
that causes the parton energy loss \cite{Bjorken:1982tu} before annihilation into a dimuon pair.
The rate of this effect is characterized by the parton energy loss in a nucleus per unit length $E'$
and effectively results in a change of the projectile parton Bjorken $x$ \cite{Garvey:2002sn}
which in turn affects the ratio in \eq{eq:dy_r}.

The data from the E866 experiment is located at somewhat lower values
of target's $x_T$ and higher values of projectile's $x_B$ with respect to the E772 data
thus falling into a region where both the shadowing and the energy loss effects become more prominent.
The analysis of Ref.\cite{KP14} indicates that the model is in a good agreement with data
at a moderate energy loss effect of order of 1 GeV/fm.
The solid curves in Fig.\ref{fig:dy} show our predictions with $E'=0.7$~GeV/fm.
Note also that the cross section ratios are taken relative to the deuterium for E772 and beryllium for E866.
For this reason the corresponding curves in Fig.\ref{fig:dy} are not identical
in the overlap region.

\section{Production of $W^\pm$ and $Z$ bosons in p+Pb collisions at LHC}
\label{sec:wz}

\begin{figure*}[t]
\begin{center}
\includegraphics[width=\textwidth]{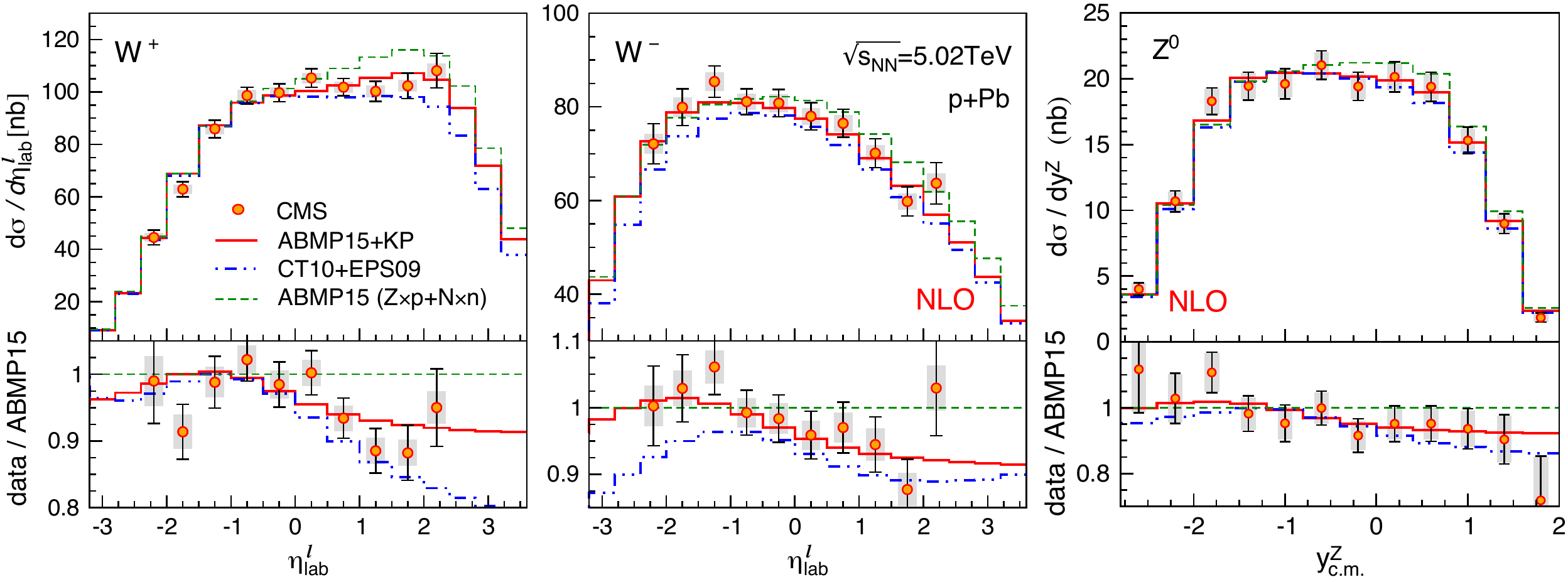}
\end{center}
\caption{%
Top panels: differential cross sections for $W^+$ (left), $W^-$ (middle), and $Z^0$ (right) production
in p+Pb collisions at $\sqrt{s_{NN}}=5.02$~TeV, as a function of (pseudo)rapidity.
The data points indicate the CMS measurements~\cite{Khachatryan:2015hha,Khachatryan:2015pzs},
while the curves show the predictions based on different models: ABMP15+KP (solid),
CT10+EPS09 (dashed-dotted), and ABMP15 without nuclear modifications (dashed).
Bottom panels: ratios of the data points and the model predictions shown in the top panels with respect
to the result obtained without nuclear modifications (ABMP15).
For more details see Ref.\cite{Ru:2016wfx}.
\label{fig:WZ}}
\end{figure*}

A study of the $W^\pm$ and $Z^0$ boson production cross sections
in p+Pb collisions with $\sqrt{s}=5.02$\,TeV at the LHC was performed in Ref.\cite{Ru:2016wfx}
in terms of the NPDF model of Ref.\cite{KP14}.
Figure~\ref{fig:WZ} shows the results of calculation of the production cross sections
as a function of the vector boson rapidity $y$
in comparison with recent CMS data~\cite{Khachatryan:2015hha,Khachatryan:2015pzs}.%
\footnote{%
Experimentally, it is easier to measure the pseudorapidity of the charged lepton
originated from the $W$ boson decay, $\eta^l$, rather than the $W$ boson rapidity $y$.
The two variables are correlated and provide similar insights on the parton distributions.}
Similar comparison with preliminary ATLAS data can be found in Ref.\cite{Ru:2016wfx}.
We found an excellent agreement between the theoretical predictions based on discussed NPDF model
and the measured observables in the entire kinematic range accessible by the experiments.
In particular, the model correctly describes the magnitude and the shape of rapidity distributions
as well as the difference between $W^+$ and $W^-$ boson distributions (flavor dependence).

This study clearly indicates the presence of nuclear modifications on
the $W/Z$ boson production cross sections
with respect to the case of p+p collisions.
The corresponding nuclear effect is illustrated in the lower panels of Fig.\ref{fig:WZ},
in which we show the ratio of the result of the full calculation to that without nuclear
corrections. In this study we use the ABMP15NNLO proton PDFs of Ref.\cite{Alekhin:2015cza}.
For comparison we also show the results obtained with CT10NLO proton PDFs \cite{Gao:2013xoa}
with phenomenological nuclear corrections of Ref.\cite{Eskola:2009uj}.

In order to illuminate the link between the nuclear correction on the rapidity distributions
and those on the PDFs, we recall that
in the leading order we have\footnote{%
This discussion aims to clarify a link between $W$ boson rapidity distributions
and PDFs. The calculation of Ref.\cite{Ru:2016wfx} was performed with the full account
of NLO corrections.}
\begin{align}
\label{lowpl}
\ud\sigma^{W^+}/\ud y &\propto \bar d_p(x_p)u_A(x_A)+u_p(x_p)\bar d_A(x_A),
\\
\label{lowmn}
\ud\sigma^{W^-}/\ud y &\propto \bar u_p(x_p)d_A(x_A)+d_p(x_p)\bar u_A(x_A),
\\
\label{x-y}
x_p &=x_0\exp(y), \quad x_A=x_0\exp(-y),
\end{align}
where $x_p$ and $x_A$ are the Bjorken variables for the proton and the nucleus, respectively,
and $x_0=M_W/\sqrt{s}$ corresponds to the central rapidity $y=0$.
At $\sqrt{s}=5.02$\ TeV we have $x_0=0.016$. Thus at $y>0$ the rapidity distributions are determined
by small-$x_A$ region in a nucleus and subject to nuclear shadowing correction. Indeed, the lower panels
in Fig.\ref{fig:WZ} show a clear suppression of p+A cross sections in this region.
Note also that in this region the production of the $W^+$ and $W^-$ is respectively driven by the
$u$ and $d$ quarks in the proton. This explains a higher magnitude of $W^+$ boson production rate and also
the different shape of $W^-$ and $W^+$ boson rapidity distribution.
The full nuclear correction on the vector boson production in p+Pb collisions is
the result of an interplay of the various mechanisms discussed in Sec.\ref{sec:npdf}.
It is interesting to note that in the backward rapidity region ($y<0$)
the off-shell correction plays an important role.

Finally, it is worth noting that the precision currently achieved by the LHC experiments
with p+Pb and Pb+Pb collisions starts to be sensitive to the predicted nuclear corrections.
A further improvement of the accuracy of future data
would be extremely valuable in this context since it could
provide a tool to disentangle the effect of
different underlying mechanisms responsible for the nuclear modifications of PDFs and to
study their flavor dependence.

\medskip
\begin{acknowledgement}
\textbf{Acknowledgements:}
I am grateful to R. Petti for fruitful collaboration on reviewed topics.
I would like to thank the Organizing Committee of Baldin ISHEPP-23 Conference for warm hospitality.
The work was supported by the Russian Science Foundation grant No.~14-22-00161.
\end{acknowledgement}


\end{document}